\documentclass[useAMS,usenatbib,usegraphicx]{mn2e}

\newcommand{\teff}{\mbox{${T}_{\rm eff}$}}

\newcommand{\msun}{\mbox{${\rm M}_{\odot}$}}

\newcommand{\simgt}{\lower.5ex\hbox{$\; \buildrel > \over \sim \;$}}
\newcommand{\simlt}{\lower.5ex\hbox{$\; \buildrel < \over \sim \;$}}
\title[Instability strips of SPB and $\beta$ Cephei stars]{Instability strips of SPB and $\beta$ Cephei stars:\\ the effect of the updated OP opacities and of the metal mixture }
\author[A. Miglio, J. Montalb\'an and M.-A. Dupret]{Andrea Miglio$^{1}$, Josefina Montalb\'an$^{1}$ 
and Marc-Antoine Dupret$^{2}$\\
$^{1}$Institut d'Astrophysique et de G\'eophysique de l'Universit\'e de Li\`ege,
All\'ee du 6 Ao\^ut, 17 B-4000 Li\`ege, Belgium\\
$^{2}$LESIA, Observatoire de Paris-Meudon, UMR 8109, 92190 Meudon, France}
\begin{document}

\date{Accepted 1988 December 15. Received 1988 December 14; in original form 1988 October 11}

\pagerange{\pageref{firstpage}--\pageref{lastpage}} \pubyear{2002}

\maketitle

\label{firstpage}

\begin{abstract}
The discovery of $\beta$ Cephei stars in low metallicity environments, as well as the difficulty in theoretically
 explaining the excitation of the pulsation modes observed in some $\beta$ Cephei and hybrid SPB-$\beta$ Cephei 
pulsators, suggest that the ``iron opacity bump'' provided by stellar models could be underestimated. We analyze
the effect of uncertainties in the opacity computations and in the solar 
metal mixture, on the excitation of pulsation modes in B-type stars. 
We carry out a pulsational stability analysis for four grids of main-sequence models with masses  between 2.5 and 12 
$\rm M_\odot$ computed with OPAL  and OP opacity tables and two different metal mixtures. 

We find that in a typical $\beta$ Cephei model the OP opacity is 25\%  larger than OPAL in the region where the driving of pulsation modes occurs. 
Furthermore, the difference in the Fe mass fraction between the two metal mixtures considered
is of the order of 20\%.  
The implication on the excitation of pulsation modes is non-negligible: the blue border of the SPB instability 
strip is displaced at higher effective temperatures, leading to a larger number of models being hybrid SPB-$\beta$~Cephei
 pulsators. Moreover, higher overtone p-modes are excited in $\beta$ Cephei models and unstable modes are found in a larger number of models for lower metallicities, in particular $\beta$ Cephei pulsations are also found in models with Z=0.01.
\end{abstract}


\section{Introduction}
At the beginning of the nineties, OPAL opacity tables \citep{RogersIglesias1992}
signified  a revolution in stellar physics. In particular, an increase of the Rosseland mean opacity ($\kappa_{\rm R}$)
as large as a factor 3 for temperatures near  3~$10^5$~K (known as ``Z--bump'') allowed to solve the long-standing
problem of B-type pulsators: $\beta$~Cep and Slowly Pulsating B-stars (SPB)  pulsate due
to $\kappa$-mechanism activated by this metal opacity bump \citep{Cox92,Kiriakidis92,Moskalik92,Dziem93b}. 
Subsequent improvements  in opacity calculations by including other iron-group elements in the metal
mixture and intermediate-coupling for Fe transitions, have led to an additional enhancement of
opacity at $\log T \sim 5.2$ and low densities, and therefore, to a  decrease of the still remaining discrepancies 
between theoretical pulsation models and observations. 
 The stability analysis of B-stars by \cite{Pamy99} for models computed with updated OPAL and OP opacity tables
\citep[Opacity Project,][ and references therein]{Seaton96} allowed to conclude that theoretical standard models
were able to explain most of the observed $\beta$~Cep and SPB stars.

However, as observational capabilities progress, B-type pulsators are now found in low metallicity environments \citep[see e.g.][and references therein]{Kolaczkowski06}, and a large number of pulsation modes is being detected in B stars, with frequency domains sometimes revealing new discrepancies between theory and observations. For instance, the two $\beta$~Cep stars 12 Lacertae  and $\nu$~Eridani 
 present low order p-modes with 
frequencies larger than those predicted by  pulsation models,  as well as high-order g-modes (SPB type oscillation) 
 \citep[][and references therein]{12Lac,nuEri}. 
In order to explain these pulsation features in $\nu$~Eri, \cite{Pamy04} proposed a local enhancement of iron
in the ``Z-bump'' region, like \cite{Cox92} suggested to overcome the difficulties to explain $\beta$~Cep 
pulsations with the first OPAL opacity tables.
  
\begin{figure*}
\begin{center}
\includegraphics[width=0.35\textwidth]{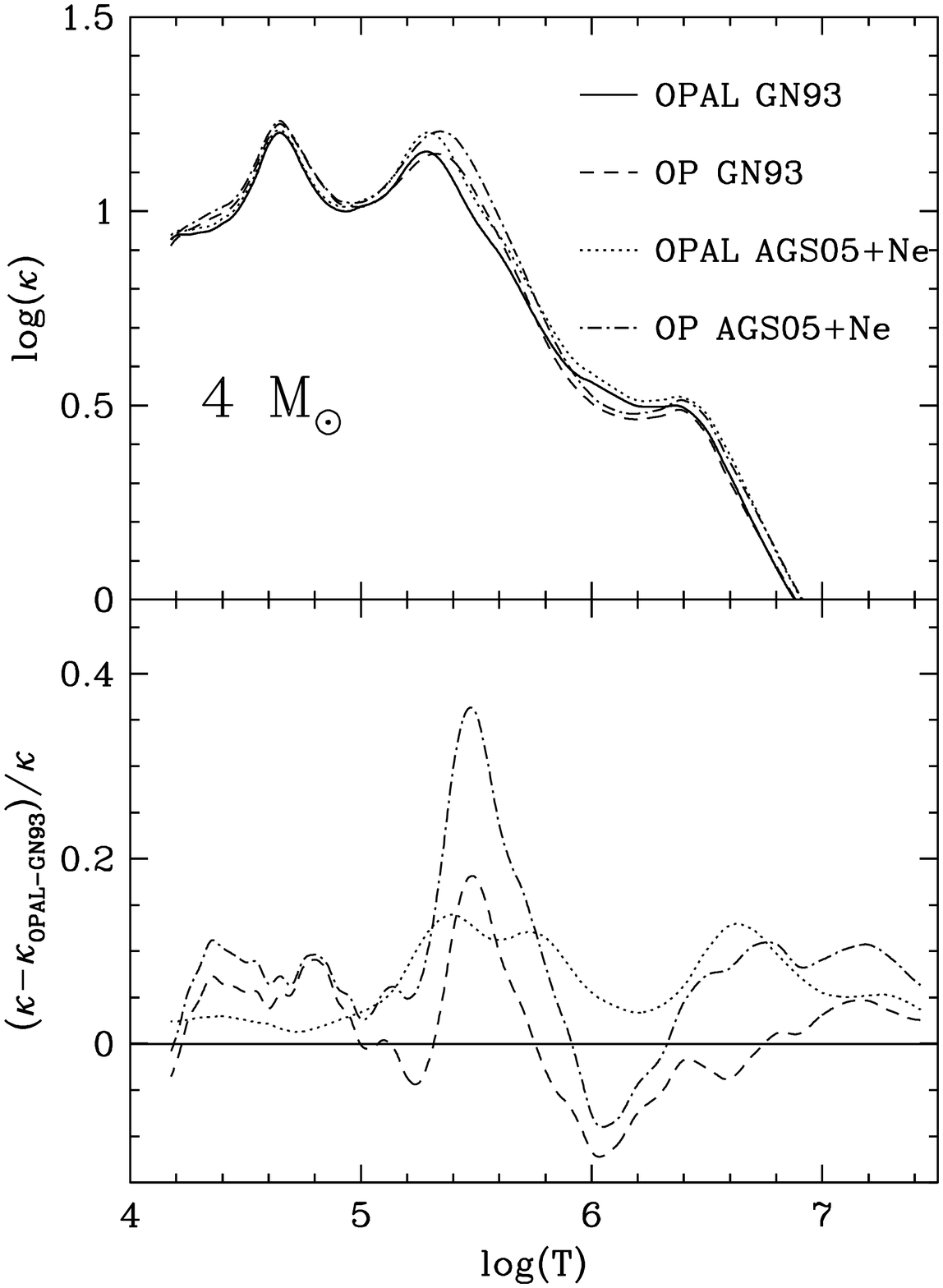}
\includegraphics[width=0.35\textwidth]{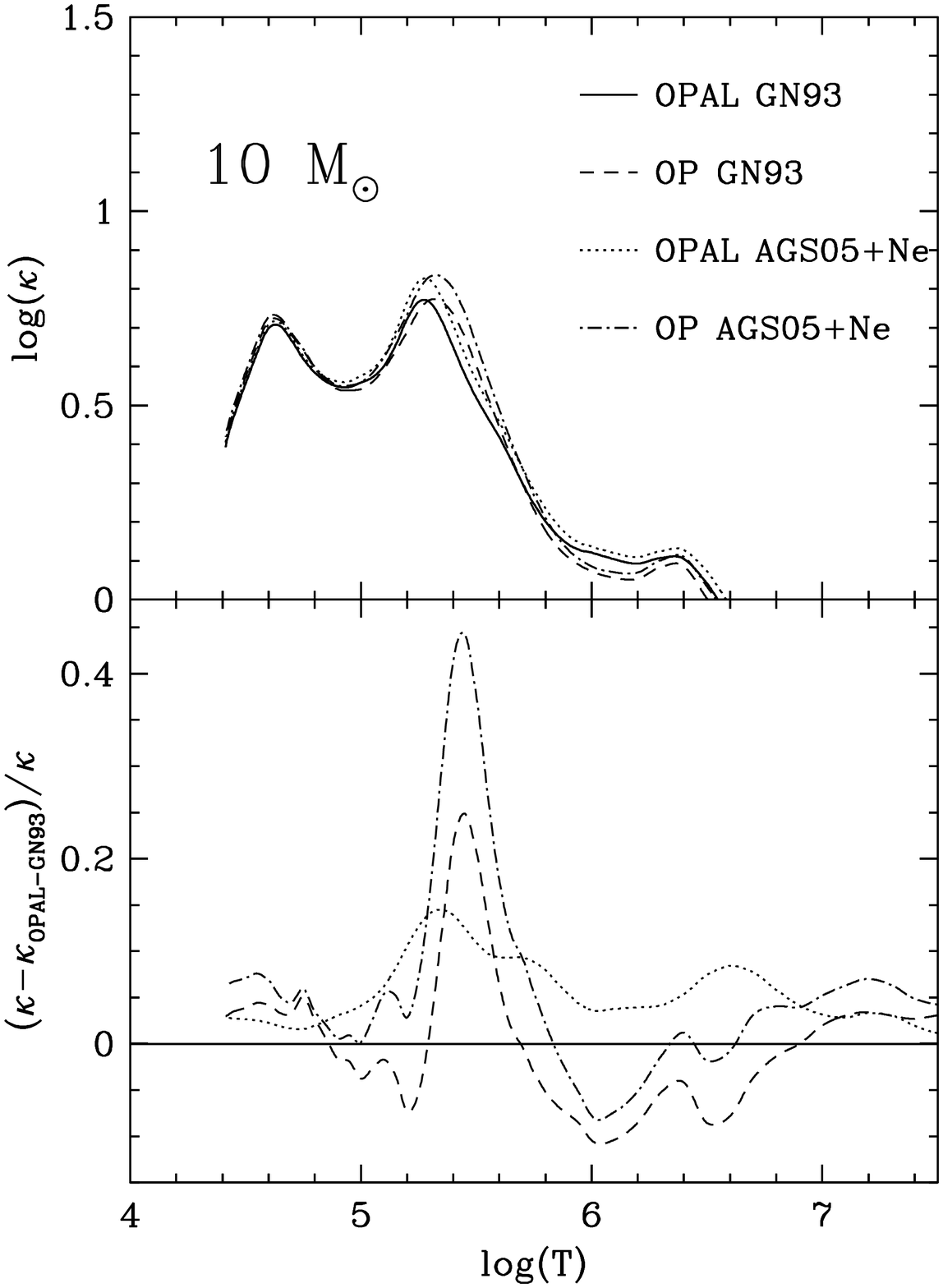}
\caption{\small Comparison between opacities computed with OPAL, OP and with different metal mixtures in the structure of a 4 and 10 $\rm M_\odot$ models on the ZAMS.}
\label{fig:opac}
\end{center}
\end{figure*}

Regardless of whether other additional physical processes must be included in the B-type stellar modeling, we propose
here to first analyze the uncertainties in the predictions of B-type pulsation models due to the 
uncertainties in the basic physical inputs in the standard model.

It is known \citep[see e.g.][]{Moskalik92} that stellar pulsation models are sensitive to $\sim$20\% changes in opacity.
Recent improvements in OP opacity computations have, on the one hand, increased the opacity in the regions where OP values 
were previously much lower than OPAL values, leading to a good agreement between both calculations.
On the other hand, the new atomic data provide a 18\% enhancement of opacity in the Z-bump region.

Besides, the chemical composition of the Sun and of B-type stars is still matter of lively debate (see, e.g. the works of \citealt{Asplund05} and \citealt{Cunha06}).
In this work we therefore investigate the impact that the uncertainties about the metal mixture have on the instability strips of B-type stars, as well as on the frequency domain of excited modes.

\section{Metal mixtures}
\label{sec-chim}
 
The  recent re-analysis of solar spectrum by \cite{Asplund05} (thereafter AGS05)  including NLTE effect as well as
 tri-dimensional model atmosphere computations,  has led to a significant 
decrease of C, N,  O  and Ne solar abundances leading to a solar metallicity 30\% smaller
than the value provided by the ``standard''  GN93 mixture \citep{Grevesse93}. 
The corresponding  decrease in solar opacity  ($\sim$~20\%) at the base of the convective envelope ruins  the good agreement
 between the standard solar model and  the seismic one.
These new CNO abundances, however,  agree with spectroscopic  measurements in B-type stars,
and solve the old discrepancy between a supermetallic Sun and the chemical composition of B-type stars
 and HII regions in the solar environment \citep[][and references therein]{Turck04, Cunha06}.
On the other hand, while the lower O abundance in AGS05 mixture decreases the opacity at the bottom of 
the convective envelope and increases the discrepancy between standard solar model and helioseismology,
an iron mass fraction 25\% larger  in the new mixture with respect to GN93 one, will favorably
affect the excitation of  $\beta$~Cep and SPB pulsation modes in early-type stars.

 Among the different solutions suggested to recover the agreement between standard model
and helioseismology, an enhancement of the Ne abundance by a factor $\sim$3.5 with 
respect to the  \cite{Asplund05} value has been proposed by \cite{Antia05} and \cite{Bahcall05}.
The increase of  Ne would compensate for the   opacity  lost  due to the 
O abundance drop. The problem is that Ne cannot be  directly measured in the solar photosphere.
\cite{Cunha06} suggested that determining non-LTE Ne abundances in a sample
of eleven B-stars members of the Orion association could help solving the problem of Ne abundance.
The studied stars span an effective temperature range from 20000~K to 29000~K. 
The result of this work is a Ne abundance in B-type stars 0.3~dex larger than in \cite{Asplund05} mixture, but CNO in
agreement with the new solar abundances.

In the following sections we will adopt  \cite{Asplund05} mixture  with A(Ne)=8.11 from \cite{Cunha06}
as the new metal mixture (AGS05+Ne) for B-type stars, and we will compare results obtained with the ``standard'' GN93.
 The re-normalization of AGS05+Ne to keep the same molecular weight results finally in an enhancement
of  $\sim20$\%  in the iron and nickel mass fractions with respect to GN93.
In  Fig.~\ref{fig:opac} the dotted curves show the relative differences in the Rosseland mean opacity between
GN93 and AGS05+Ne  (for the same metal mass fraction Z=0.02) 
for the structure of a 4 and a 10 \msun\ models. The effect of the
enhancement of Fe and Ni  is to increase the height and the width of the  Z-bump of the Rosseland mean opacity ($\kappa_{\rm R}$).
The consequences on the excitation of pulsation models are described in Sec.~\ref{sec-stability}.

\section{Opacities: OPAL versus OP}
\label{sec-op}
OP opacity tables  have  been recently updated including data for the inner-shell transitions,
 inter-combination lines and improvements in photoionization cross-sections \citep{Seaton05, Badnell05}.
These changes result in an enhancement of $\kappa_{\rm R}$ at high
density and temperatures, and an increase of 18\% of opacity in the Z-bump due to the new
Fe atomic data.
The new OP opacity tables are much closer to the OPAL ones than they were previously \citep{Seaton94}.
In particular, the differences at high temperature and high density have almost disappeared.  
Some differences remain, however, at low temperatures ($\log T < 5.5$): 
the OP Z-bump in $\kappa_{\rm R}$  presents a hot wing slightly larger than the OPAL one.
The comparisons between OP and OPAL by \cite{Badnell05} at fixed $\log R$ ($R=\rho/T_6^3$) 
show that for high densities 
the differences are of the order of 5--10\%, but at low density ($\log R \sim -4$), 
differences up to 30\% appear due to the differences in the high-temperature wing of the Z-bump.
 
The previous OP tables \citep{Seaton94} already presented a  Z-bump  in $\kappa_{\rm R}$ 
shifted by 15000--20000~K to higher temperatures compared with OPAL.
 The effect of this difference on the instability strip
of $\beta$~Cep and SPB stars was studied by \cite{Pamy99}.
In that study, however, OP and OPAL opacity tables had also small  differences in heavy elements abundances:
Ni ~ 5\% more abundant in OP than in OPAL, and  Ne and Fe $\sim 2-3$\% more abundant in OP.

 Ni and Fe are the main contributors  to the Z-bump, and 
 Ni contributes to the Z-bump in $\kappa_{\rm R}$ at higher 
temperatures than Fe \citep[see e.g.][]{jeffery06}. In order to separate metal mixture effects from
opacity calculation ones, we obtained OP opacity tables from the OPserver\footnote{\texttt http://vizier.u-strasbg.fr/topbase/op.html} 
for the same GN93 and AGS05+Ne metal mixture as OPAL. The abundances of the four elements that are not included in OP 
(P, Cl, K, Ti) are redistributed among the other 15 elements. Doing so, the abundance differences between  
OP and OPAL opacity table computation are at maximum 0.06\%.

In Fig~\ref{fig:opac} we show the difference in opacity for the internal structure of two
stellar models with the typical masses of an SPB (4~\msun) and a $\beta$~Cep (10~\msun). We see that 
for the lower-mass model the opacity differences in the Z-bump region are of the order of 20\%, and even larger for the 10~\msun\ model.

\section{Stellar models}

Stellar models have been computed with the code CLES (Code Li\'egeois d'Evolution Stellaire, \citealt{Scuflaire05}).
The main physical inputs are:
OPAL2001 equation of state \citep{Rogers02} and  \cite{Caughlan88} nuclear reaction rates  with \cite{Formicola04}
for the $^{14}$N(p,$\gamma$)$^{15}$O cross-section. Convective transport is treated by using the
classical Mixing Length Theory of convection \citep{Bohm58},  and a convective overshooting
parameter of 0.2 pressure scale height was assumed in all the models.
For the chemical composition we have considered: GN93 and AGS05+A(Ne)=8.11.  
We have computed models with: {\it i)} OPAL opacity tables with GN93 and {\it ii)} AGS05+Ne chemical composition, then models with 
{\it iii)} OP opacity tables assuming GN93 and {\it iv)} AGS05+Ne mixtures.
All the opacity tables are completed at $\log T < 4.1$  
with the corresponding GN93 and AGS05 low temperature tables by \citet{Ferguson05}.
The masses considered span from 2.5 to 12~\msun, and the chemical composition considered are: 
$X=0.70$ for the hydrogen mass fraction and three different metal mass fractions, $Z=0.02$, 0.01 and 0.005.
For all the models the evolution was followed from the Pre-Main Sequence.

\begin{figure*}
\begin{center}
\includegraphics[width=0.85\textwidth,angle=0]{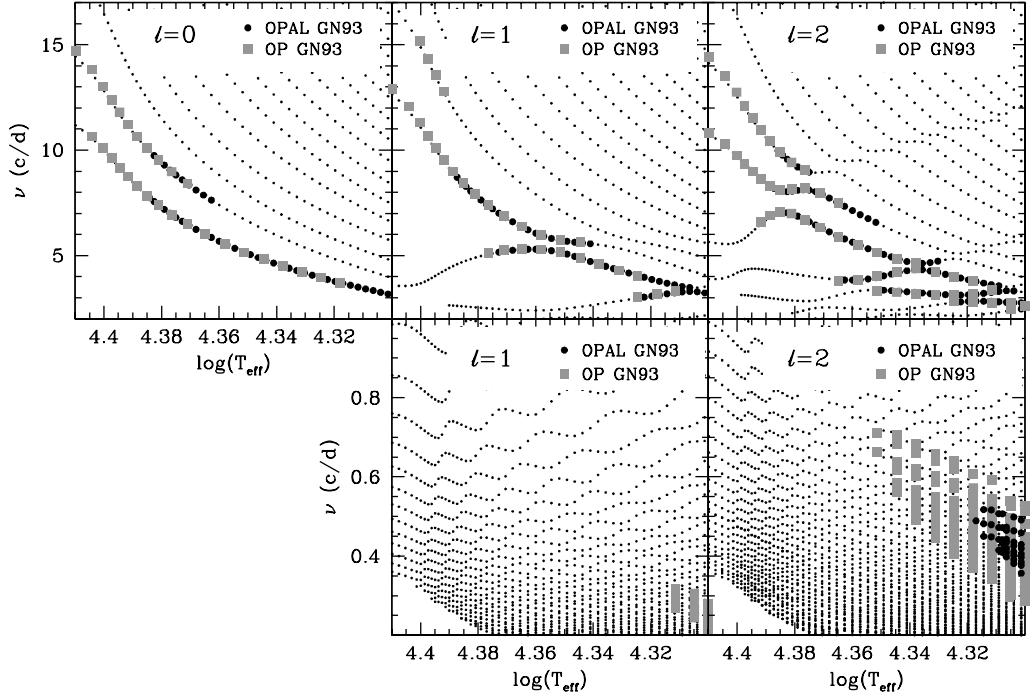}
\vspace*{-5.cm}
\caption{\small Frequencies of pulsation modes as a function of $\log{\teff}$ for main-sequence models of a 10 $\rm M_\odot$  Z=0.02 star. Unstable modes are described by large black dots and gray squares when models are computed, respectively, with OPAL-GN93 and OP-GN93 opacity tables. Lower panels represent high-order g-modes.}
\label{fig:opopal}
\end{center}
\end{figure*}

\begin{figure*}
\vspace{-1cm}
\begin{center}
\includegraphics[width=0.85\textwidth,angle=0]{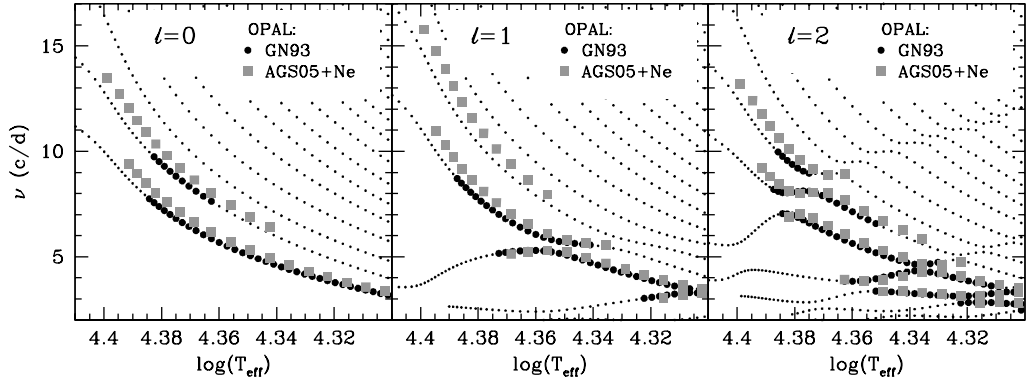}
\vspace*{-9.0cm}
\caption{\small As in Fig. \ref{fig:opopal} but comparing models computed with OPAL-GN93 and OP-AGS05+Ne opacity tables.}
\label{fig:gn93-a05}
\end{center}
\end{figure*}


\section{Stability analysis}
\label{sec-stability}
Since oscillations in $\beta$~Cep and SPB stars are driven by the $\kappa$-mechanism acting in the Z opacity bump, 
the location of the instability strip in the HR diagram and the frequency of the excited modes are determined by the properties 
of the metal opacity bump. In models with the new OP opacity tables, the driving region, that is, the region where
$\kappa_{\rm T}=\left(\partial\log{\kappa_{\rm R}}/\partial \log{T}\right)_\rho$ increases outwards, is found deeper in the star with respect to
the models computed with  OPAL. As a consequence, we expect hotter B-type pulsators with OP models compared
to OPAL ones.

We perform stability analyses of  main-sequence models from our grid using the non-adiabatic code MAD \citep{Dupret03}.
 Fig.~\ref{fig:opopal} shows different effects of OP and OPAL on the p-modes excitation (upper panels) and on the high-order g-modes excitation
(lower panels) for main sequence models of a 10~\msun\ star with initial chemical composition given by 
$X=0.70$, $Z=0.02$ and GN93 metal mixture.
For p-modes (or low-order g-modes), we see that in the common instability region, OPAL and  OP predict the
same excited modes, and their frequencies are  not affected  by opacity differences. 
There is, however, a significant effect on the 
hottest unstable models. While OPAL models with excited radial modes are expected only for $X_{\rm c} < 0.5$,
ZAMS OP models have $\ell=0$ modes excited.

\begin{figure}
\begin{center}
\resizebox{1.05\hsize}{!}{\includegraphics{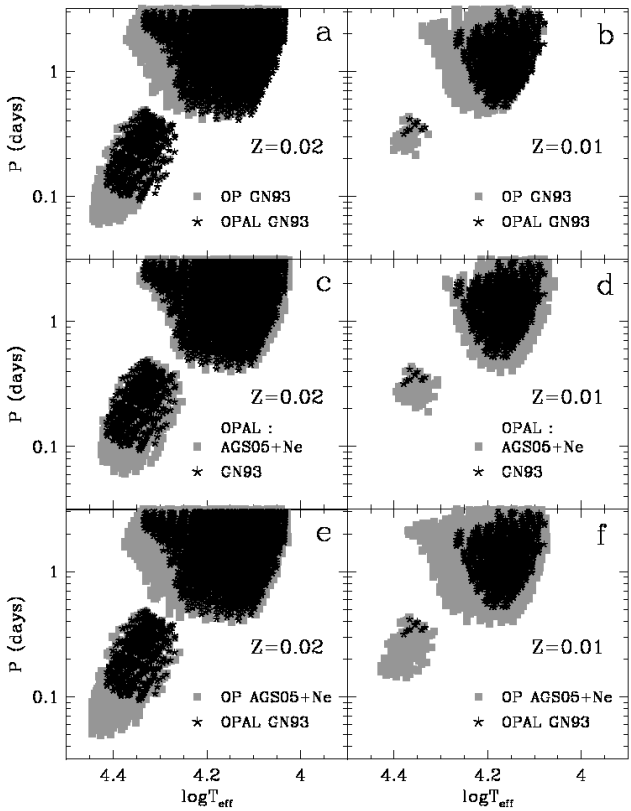}}
\vspace{-0.8cm}
\caption{\small Instability strips represented in a $\log{\teff}$-$\log{P}$ diagram. In each panel, the two regions of unstable modes represent $\beta$ Cep- and SPB-type pulsations.}
\label{fig:BI}
\end{center}
\end{figure}

The differences between OP and OPAL are even more pronounced for high-order g-modes. There are many more excited modes for
OP models than for the OPAL ones. For instance, for a 10~\msun\ star, the OP model with $X_{\rm c}=0.2$ presents
excited modes of $\ell=2$ going from $g_{16}$ to $g_{21}$, and the TAMS model has $\ell=1$, $g_{31}$--$g_{43}$ and $\ell=2$, 
$g_{28}$--$g_{53}$ excited. In OPAL models, instead, no $\ell=1$ high-order g-mode is excited; the first excited $\ell=2$ g-modes appear
at  $X_{\rm c}=0.1$, and in the TAMS model the order of the excited g-modes goes from $n=34$ to $n=48$.
The \teff\ domain for which we find excited SPB-modes is therefore $\sim$3000~K larger
than for OPAL models. As a consequence,
the number of expected  hybrid $\beta$~Cep--SPB objects is also larger for OP models  
(see right panels in Fig.~\ref{fig:opopal} and Fig~\ref{fig:BI}a and c).
These results are similar to those obtained by \cite{Pamy99} with the previous OP opacity tables. In fact
the higher Fe and Ni abundances in the original OP mixture, are balanced now by a 18\% higher 
opacity at the Z-bump.
\cite{Pamy99} noted that the shift of the Z-bump towards higher temperatures leads to 
 an increase of the  $\beta$~Cephei and SPB  instability domains towards higher \teff\ and luminosity
with respect to OPAL models.  
These results, however, have not been taken into consideration to explain the high order g-modes
detected in the $\beta$~Cep stars 12~Lac and $\nu$~Eri.

The instability strips for OP and OPAL models, with  GN93 mixture and metal contents Z=0.02 and Z=0.01 are shown
in Fig.~\ref{fig:BI} (panels a and b). We see that the impact of OP is more important for lower metallicity. In fact, the
expected OP $\beta$~Cep pulsators in the LMC (Z=0.01) would be more than twice larger than estimated  from OPAL stellar models.
 
The Fe-mass fraction enhancement in the AGS05+Ne mixture, compared with GN93, has the main effect of extending towards higher overtones the 
range of excited frequencies. 
Fig.~\ref{fig:gn93-a05} shows the $\ell=0$, 1 and 2 excited p-modes for 10~\msun\ main sequence models 
calculated with OPAL opacity tables for the two different metal mixtures. 
AGS05+Ne models also provides slightly wider instability bands,
and this effect increases as metallicity decreases (see Fig.~\ref{fig:BI}b and d), thus, the expected $\beta$~Cep pulsators for LMC 
predicted with AGS05+Ne would be more than three times larger than with GN93.

Finally, if we consider the combined effect of OP opacities and AGS05+Ne metal mixture, we see that the large difference in $\kappa_{\rm R}$ described in Fig. \ref{fig:opac} is reflected in wider instability strips, in the excitation of higher overtones in $\beta$~Cep models and in a significantly larger number of $\beta$~Cep pulsators for Z=0.01 (see Fig. \ref{fig:BI}e and f).

Computations for the lower metallicity corresponding to SMC (Z=0.005), show that none of the different OP/OPAL and GN93/AGS05+Ne 
evolutionary tracks for masses up to 12\msun\ predicts $\beta$~Cep pulsators, whereas we find SPB-type modes excited when considering OP with AGS05+Ne.

\section{Conclusions}
We have shown that the current uncertainties on opacity calculations and on the solar metal mixture adopted in standard stellar models, have a considerable effect on the excitation of pulsations in $\beta$~Cep and SPB stars. Compared to models computed with OPAL opacities and GN93 metal mixture, we find that with OP opacities high-order g-modes are predicted to be excited in hotter stars and that higher overtones are excited if the AGS05+Ne metal mixture is considered. For low metallicities, such as Z=0.01, a larger number of models with excited modes is also found. These findings could help solving, at least partly, the discrepancies between theoretical predictions and observations of $\beta$ Cep in low-metallicity environments and of a large domain of excited modes detected in some $\beta$~Cep and hybrid SPB-$\beta$~Cep pulsators.
The detailed modelling of single stars, such as $\nu$ Eri is, however, beyond the scope of this letter and will be addressed in a forthcoming paper.
\section*{Acknowledgments}
A.M. and J.M  acknowledge financial support from the Prodex 8 COROT (C90199).
\vspace{-0.2cm}
\bibliographystyle{mn2e}
\small
\bibliography{opacity}
\label{lastpage}
\end{document}